\documentclass[epj]{svjour}
\usepackage[normalem]{ulem}
\usepackage{graphicx}
\usepackage{color}

\usepackage{xcolor}
\usepackage{subfigure}
\def\scr{\rm\scriptscriptstyle }
\usepackage{amssymb}
\usepackage{subfigure}
\usepackage{amsmath}
\usepackage[mathscr]{eucal}
\usepackage{lipsum} 

\def\scr{\rm\scriptscriptstyle }

\begin{document}
\title{Semi-classical approaches to heavy-ion reactions: fusion, rainbow, and glory}
\author{L.F. Canto\inst{1} \and K. Hagino\inst{2} \and M. Ueda\inst{3}
}                     
%
%
\institute{
Instituto de F\'{i}sica, Universidade Federal do Rio de Janeiro, 
CP 68528, Rio de Janeiro, Brazil 
\and
Department of Physics, Kyoto Unviersity, Kyoto 606-8502, Japan
\and
National Institute of Technology, Akita College, Akita 011-8511, Japan
}
\date{Received: date / Revised version: date}
%
\abstract{
A semi-classical approximation has been a powerful tool in understanding 
the dynamics of 
low-energy heavy-ion reactions. 
Here we discuss two topics in this regard, for which Mahir Hussein was a 
world leading pioneer. 
The first topic 
is heavy-ion fusion reactions of neutron-rich nuclei, in which 
the breakup process of the projectile nucleus plays a crucial role. The second is 
rainbow and glory scattering, for which characteristic oscillatory patterns 
in differential cross sections 
can be well understood in terms of intereferences among 
several semi-classical trajectories. 
} 
\maketitle
\section{Introduction}

The semi-classical approximation provides an intuitive view of various 
quantum mechanical phenomena in terms of classical concepts, such as a 
trajectory of a particle. This approximation works well when a variation of a 
potential within a wave length is negligibly small. The condition is satisfied when 
the energy and/or the mass of a particle is large. 
This is well fulfilled in 
heavy-ion reactions \cite{Bri85,CaH13,FrL96}, 
for which a reduced mass for the relative motion between 
nuclei is in general large. 
As a matter of fact, angular distributions of elastic scattering 
can often 
be interpreted using classical trajectories. Also, a semi-classical 
coupled-channels method, in which quantal coupled-channels equations are 
solved along a classical trajectory, has been developed for inelastic 
scattering \cite{BrW04}. Moreover, the WKB formula for a penetrability 
has provided a convenient reference for heavy-ion fusion reactions at 
energies close to the Coulomb barrier. 

Mahir Hussein was an expert of the semi-classical approximation 
in the context of heavy-ion reactions, and carried out many pioneering works, 
as is well summarized in his textbook on nuclear reactions written with 
one of us (L.F.C.) \cite{CaH13}. 
In this paper, we shall discuss two topics among them. First is heavy-ion fusion 
reactions of neutron-rich nuclei. Here, the weakly bound nature of a projectile 
nucleus leads to complex features in the fusion dynamics. While a halo structure of a 
projectile nucleus naturally lowers the Coulomb barrier, the reaction dynamics 
is much more complicated due to the breakup process \cite{CGD06,CGD15}. 
Hussein was the first who discussed the role of breakup in fusion 
of weakly bound nuclei \cite{HPC92}. 
The second topic which we discuss in this paper is heavy-ion elastic scattering. 
Differential cross sections often exhibit characteristic oscillations. In the 
semi-classical approximation, such oscillations can be naturally interpreted 
as intereferences between several trajectories, such as intereferenes between 
a near-side and a far-side components of scattering amplitude \cite{HuM84}, 
and intereferences between an internal wave and a barrier wave \cite{BrT77a}.  
Among them, rainbow scattering is particularly important, as it probes 
a relatively inner region of an optical potential and 
thus it can be used to constrain 
the depth of a potential \cite{KOB07}. 
Glory scattering is another interesting phenomenon in 
which many classical trajectories coherently contribute to cross sections. 
In his own terms, Hussein mentioned ``These effects (nuclear rainbow and glory 
scattering) are also common in atomic and molecular scattering, and 
have been reviewed extensively in the literature. Of course the commonest of all 
is the atmospheric rainbow and glory, a beautiful colorful dance 
of light and water'' (at the workshop on occasion of Noboru Takigawa's 60th 
birthday, November 2003, Sendai, Japan). 
Here in this paper we 
shall discuss the novel concept of glory in the shadow of rainbow, 
introduced by Hussein and his collaborators in Ref. \cite{UPH98,UPH99}. 

\section{Fusion of weakly bound nuclei}

\subsection{A two-neutron halo nucleus $^{11}$Li}

Nuclei far from the stability line are characterized by an extended 
density distribution due to weakly bound valence nucleons. 
Among such neutron-rich nuclei, the $^{11}$Li nucleus has been one of the 
most well studied nucleus since the discovery 
of its halo structure \cite{THH85}. 
The two-neutron separation energy of this nucleus is indeed small, 
$S_{2n}=378\pm5$ keV \cite{BAG08}. A strong low energy peak 
has been experimentally found in the Coulomb breakup 
spectrum \cite{NVS06}, which 
is consistent with the small separation energy \cite{BBH91}. 
The low energy peak is due to the electric dipole (E1) excitation and 
thus has been referred to as a soft dipole mode. 
If 
one employs a three-body model with $^9$Li+$n$+$n$ for the $^{11}$Li, 
the operator 
for the electric dipole excitation is proportional to $\vec{r}_1+\vec{r}_2$, where 
$\vec{r}_i$ is the coordinate of the $i$-th neutron measured 
from the core nucleus $^9$Li \cite{BeE91,HaS05,SaH15}. 
Using the cluster sum rule, the total 
E1 strength is thus proportional to the ground state expectation value of 
$\vec{R}^2$, where $\vec{R}=(\vec{r}_1+\vec{r}_2)/2$ is the center of mass 
of the valence neutrons with respect to the core nucleus. 
This implies that if one somehow supplies information on the distance between 
the two valence neutrons, $\vec{r}_{nn}=\vec{r}_1-\vec{r}_2$, one can combine those 
information to reconstruct the 
three-body geometry of the $^{11}$Li nucleus. 
This was done by Bertulani and Hussein, who 
used a HBT-type analysis of the two valence-halo 
particles correlation to extract the opening angle of the valence neutrons in 
$^{11}$Li to be $\theta_{nn}=66^{+22}_{-18}$ degrees \cite{BeH07}. See also 
Ref. \cite{HaS07}, which used the matter radius to estimate $r_{nn}$ 
and obtained $\theta_{nn}=56.2^{+17.8}_{-21.3}$ degrees. 
These values are consistent with each other and both are smaller 
than the value of the uncorrelated case, that is, 
90 degrees, implying the existence of the dineutron correlation 
\cite{HaS05,MMS05,Mat06} in $^{11}$Li. 

\subsection{Sub-barrier fusion of $^{11}$Li} 

Heavy-ion fusion reactions take place by quantum tunneling at energies 
below the Coulomb barrier, and they are 
sensitive to details of nuclear structure of colliding nuclei. 
In particular, it has been well known that collective excitations 
of the colliding nuclei significantly enhance fusion cross sections at subbarrier 
energies \cite{BaT98,DHR98,HaT12,BEJ14,MoS17}. 
A natural question is then how fusion cross sections are affected when a 
neutron-rich nucleus is used as a projectile. There are several aspects which 
one has to take into account. Those include: 

\begin{itemize}
\item the extended density distribution of the projectile. This lowers the 
Coulomb barrier, enhancing 
fusion cross sections \cite{TaS91}. 
\item the soft dipole excitation. Even though it may not carry a large  
collectivity \cite{SGT95}, 
couplings to continuum may in general enhance fusion cross sections. 

\item the breakup process. It may hinder fusion cross sections since the 
lowering of the Coulomb barrier disappears. At the same time, it may also 
enhance fusion cross 
sections if couplings to a breakup channel 
dynamically lowers the Coulomb barrier \cite{DaV94,HVD00}.

\item the transfer processes. For neutron-rich nuclei, a transfer $Q$-value is 
likely positive, which may significantly influence fusion reactions \cite{CCS18}. 
It is still a challenge to take into account simultaneously both 
the transfer and the 
breakup processes in a theoretical calculation \cite{IYN06a}. 
\end{itemize}

Among these effects, in this paper 
we particularly focus on the effect of breakup on heavy-ion fusion 
reactions. This problem was discussed for the first time by Hussein et al. \cite{HPC92}. 
Couplings to a breakup channel yields a dynamical polarization potential, $V_{\scr DPP}$, 
for the entrance channel. 
By taking into account the imaginary part of the dynamical polarization potential, 
Hussein {\it et al}. estimated fusion cross sections in the presence of a breakup channel 
as, 
\begin{multline}
\sigma_{\rm fus}(E)=\frac{\pi}{k^2}\sum_l(2l+1)\,\Bigg[ \frac{1}{2} \Big( T_l(E+F) + T_l(E-F)\Big) \Bigg] \\
\times \mathcal{P}^{\scr S}_l(E),
\label{bufus}
\end{multline}
where $E$ is the incident energy in the center of mass frame, $k=\sqrt{2\mu E/\hbar^2}$ is the wave number 
for the relative motion, with $\mu$ being the reduced mass, and $F$ is the strength of the coupling between 
the elastic channel and the soft dipole mode, taken at the barrier radius. The contribution from each partial
wave involves the product of two factors. The first is the sum of the transmission probabilities through the barrier 
of the $l$-dependent potential for the collision energies $E\pm F$. The energy shifts, $\pm F$, account for the 
effects of couplings with the soft dipole mode in an approximate way~\cite{LiR84,DLW83}.
The second term is the probability of surviving the prompt breakup process. It is given by
\begin{equation}
\mathcal{P}^{\scr S}_l(E) = 1-\mathcal{P}_l^{\rm bu}(E),
\label{Psurv}
\end{equation}
with  $\mathcal{P}_l^{\rm bu}(E)$ being the breakup probability, estimated semi-classically as, 
\begin{equation}
\mathcal{P}_l^{\rm bu}(E)=1-\exp\left[2\int^\infty_{r_{0l}}W_{\scr DPP}(r)\frac{dr}{\hbar v(r)}\right] .
\end{equation}
Above,  $r_{0l}$ is the outermost classical turning point for $l$, $v(r)$ is the local velocity, and $W_{\scr DPP}$ is the 
imaginary part of the dynamical polarization potential. Notice that the exponent in this equation is a half of that in 
Ref. \cite{HPC92}, by taking into account only the incoming part of the trajectory \cite{TKS93}. Since $W_{\scr DPP}$ is 
negative, in this approach fusion cross sections are suppressed due to the breakup.

\begin{figure}[tb]
\begin{center}
  \includegraphics*[width=8cm]{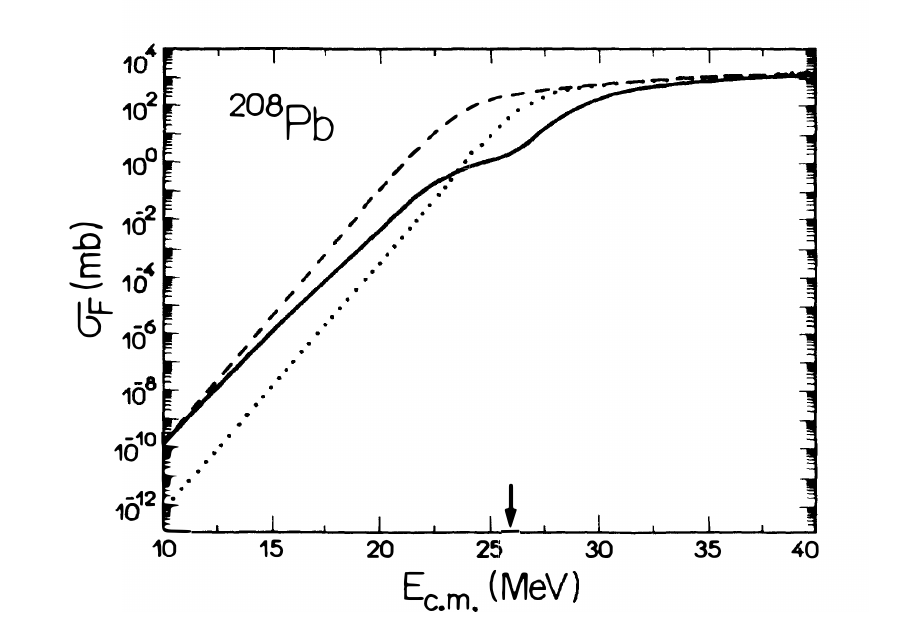}
\caption{Fusion cross sections for the $^{11}$Li+$^{208}$Pb system. 
The dotted line is obtained with a single-channel calculation, while 
the dashed line is obtained by taking into account couplings to a soft 
dipole model excitation in $^{11}$Li. The solid line takes into account the 
breakup process in the semi-classical approximation, in addition to the couplings 
to the soft dipole excitation. Taken from Ref. \cite{HPC92}. }
\end{center}
\end{figure}

Fig. 1 shows fusion cross sections for $^{11}$Li+$^{208}$Pb, within different approximations. The solid line represents the cross section of
Eq.~(\ref{bufus}), which takes into account both couplings to the soft dipole mode and the influence of prompt breakup. The dashed line 
corresponds to the cross section taking into account couplings to a soft dipole excitation in $^{11}$Li, but not survival probabilities 
($F\ne 0$ but $\mathcal{P}^{\scr S}_l(E)= 1$). Finally, the dotted line corresponds to results of a one-channel calculation, which neglects all 
coupling effects ($F=0$ and $\mathcal{P}^{\scr S}_l(E)= 1$). Comparing the solid and the dotted lines, one concludes that the overall
effect of the couplings is suppression of fusion above the Coulomb barrier and enhancement at sub-barrier energies. One can see also that,
the cross section of Eq.~(\ref{bufus}) converges to the dashed line as the energy decreases well bellow the Coulomb barrier, meaning that
in this energy limit the effects of prompt breakup become negligible, while the influence of the soft dipole mode remains.\\

Dasso and Vitturi~\cite{DaV94} proposed a different approach to estimate breakup effects
in $^{11}$Li+$^{208}$Pb fusion. They performed schematic coupled channel calculations
involving two $^{11}$Li channels, corresponding to the elastic channel and the low-lying
soft dipole mode, and a third channel for $^9$Li, associated with the  breakup process.  
They found that the couplings with the soft mode makes the fusion cross section much
larger, at all collision energies. Further, they found that the inclusion of the $^9$Li channel
in the coupled equations makes the cross section still larger. 

\bigskip

These early calculations were based on very drastic approximations. 
Dasso and Vitturi treated the breakup channel only schematically. On the 
other hand, Hussein {\it et al.} evaluated the imaginary part of the polarization 
potential with several approximations, and completely neglected its real part. 
In reality, one has to take into account both the real and the imaginary parts of 
the dynamical polarization potential. The real part of the dynamical polarization potential is 
attractive at energies below the Coulomb barrier \cite{HPC93}, and the coupling to the breakup 
process may lead to an enhancement. More quantitative theories were developed along the
last three decades, including realistic quantum mechanical calculations, 
based on the continuum discretized coupled channel (CDCC) method~\cite{SYK83,SYK86,AIK87}. 
A summary of these theories is presented below.

\subsection{Further developments in the treatment of breakup in fusion}

In collisions of neutron-halo projectiles, the experimental fusion cross section has contributions from captures of 
the whole projectile, and of the charged core, produced in prompt breakup. The two processes are 
experimentally indistinguishable. However, the latter contribution is expected to be small, owing to the higher
Coulomb barrier for the charged fragment and also to its lower kinetic energy. This justifies the neglect of  $^9$Li 
fusion in Ref.~\cite{HPC92}. \\

A different situation occurs in collisions of weakly bound projectiles that break up into two charged fragments.
Some examples are the stable $^6$Li ($^4$He\,+\,$^2$H) and $^7$Li ($^4$He\,+\,$^3$H) nuclei, and the unstable 
$^8$B ($^7$Be\,+\,$p$). 
In such cases, the fusion of the whole projectile, known as complete fusion (CF), may, in principle, be experimentally
distinguished from the fusion of one of the breakup fragments, known as incomplete fusion (ICF). Some experiments
can determine also individual cross sections for the captures of the two breakup fragments. Then, the situation calls
for more powerful theoretical models, that can predict cross sections for CF and also individual ICF cross sections for
the two fragments, denoted by ICF$_{\scr 1}$ and ICF$_{\scr 2}$. \\

\begin{figure}[ptb]
\begin{center}
\includegraphics*[width=9cm]{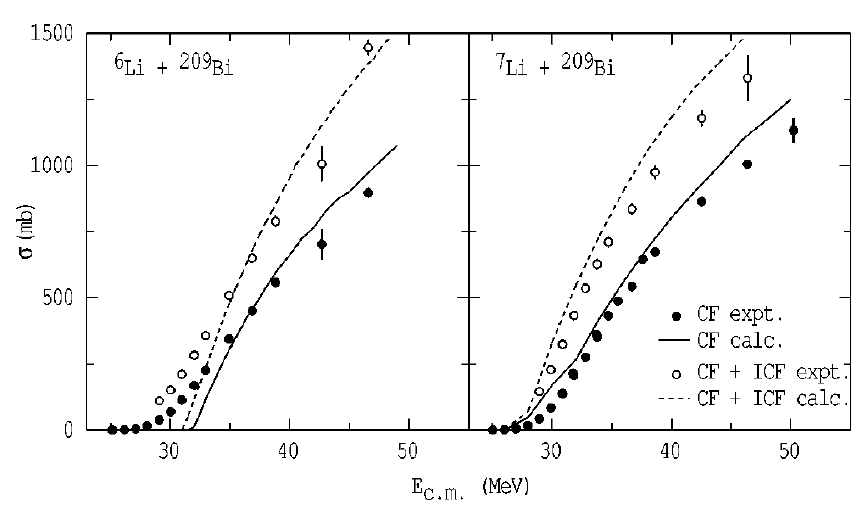}
\end{center}
\caption{ CF and TF cross sections in the $^{6,7}$Li+$^{209}$Bi collisions of the classical model, in comparison with the data. 
The calculations and the data are from the work of Dasgupta {\it et al.}~\cite{DHH02}. }%
\label{sigF-class}%
\end{figure}

The first theory to evaluate CF and ICF cross sections was introduced in Ref.~\cite{DHH02}, which reported also measurements 
of CF and ICF cross sections in $^{6,7}$Li + $^{209}$Bi collisions. A detailed presentation of the theory can be found in 
Ref.~\cite{HDH04}. It treats the collision by a classical three-body model, describing the motion of the target (T) and the two 
clusters of the projectile ($c_{\scr 1}$ and $c_{\scr 2}$) on the x-y plane. The time evolution of the system is determined by the 
Hamiltonian,
\begin{multline}
H=\frac{\vec{p}_{\scr T}^2}{2m_{\scr T}}+\frac{\vec{p}_{\scr 1}^2}{2m_{\scr 1}}
+\frac{\vec{p}_{\scr 2}^2}{2m_{\scr 2}}+ V_{\scr 12}\left( \vec{r}_{\scr 1}-
\vec{r}_{\scr 2} \right)\\
+ V_{{\scr 1T}}\left( \vec{r}_{\scr 1}-\vec{r}_{\scr T} \right) +
V_{{\scr 2T}}\left( \vec{r}_{\scr 2}-\vec{r}_{\scr T} \right).
\label{Hclass}
\end{multline}
Above, $\vec{r}_{\scr T}$, $\vec{r}_{\scr 1}$ and $\vec{r}_{\scr 2}$ are vectors in the x-y plane, representing respectively the target, 
fragment $c_1$ and fragment $c_2$, and $\vec{p}_{\scr T}$, $\vec{p}_{\scr 1}$ and $\vec{p}_{\scr 2}$ are the corresponding momenta.
The three potentials represent the interactions between the particles of the model ($T, c_{\scr 1}$ and $c_{\scr 2}$). The time 
evolution of the system begins with the projectile far apart from the target, where 
its interactions with the fragments depend
only on the projectile-target relative vector, $\vec{R} = \vec{r}_{\scr P} - \vec{r}_{\scr T}$, with $\vec{r}_{\scr P} = (m_{\scr 1} \vec{r}_{\scr 1} 
+ m_{\scr 2} \vec{r}_{\scr 2})/ (m_{\scr 1} + m_{\scr 2})$. The initial projectile-target momentum is given by the collision energy, 
whereas the initial values of $\vec{r}_{\scr 1}, \vec{p}_{\scr 1}, \vec{r}_{\scr 2}, \vec{p}_{\scr 2}$ are chosen randomly, from a distribution
of positions and momenta given by the ground state (g.s.) 
wave function of the projectile. The calculations are
performed for a mesh of impact parameters and the CF, ICF$_{\scr 1}$ and ICF$_{\scr 2}$ are determined from the final states 
of the three-body system. This model was used to predict CF and TF (${\rm CF} + {\rm ICF}_{\scr 1} + {\rm ICF}_{\scr 2}$) cross 
sections. The results for the $^{6,7}$Li + $^{209}$Bi systems are shown in Fig.~\ref{sigF-class}, in comparison with the data. 
Despite the simplicity of the model, it gives a reasonable account of the data at above-barrier energies. Of course, a classical 
model cannot describe sub-barrier fusion.\\

Diaz-Torres {\it et al.}~\cite{DHT07,Dia10} developed a three-dimensional version of the classical model, in which the breakup of 
the projectile was treated as a stochastic process, based on a breakup function determined from sub-barrier breakup 
measurements~\cite{HDF02}, or from CDCC calculations. This model was implemented in the PLATYPUS computer code, available in 
the literature~\cite{Dia11}. It was used to evaluate CF and ICF cross sections in the $^9$Be - $^{208}$Pb \cite{DHT07} and
$^{6}$Li - $^{209}$Bi \cite{DiQ18} collisions. Their results were compared with the data of Refs.~\cite{DHB99} and \cite{DHH02,DGH04},
respectively. The theoretical cross sections 
were shown to be in qualitative agreement with the above-barrier data.\\

Although the classical models give reasonable descriptions of the CF and ICF cross sections above the Coulomb barrier, they do
not include quantum mechanical effects, like barrier tunnelling, which are essential at sub-barrier energies. The situation is improved 
in the semiclassical model (see e.g. Ref.\cite{AlW75}). The model is based on the same Hamiltonian of Eq.~(\ref{Hclass}), but
in the c.m. of the projectile-target system, and expressed in terms of the 
vectors $\vec{R}$ and $\vec{r} = \vec{r}_{\scr 1} - \vec{r}_{\scr 2}$.
That is,
\begin{equation}
H=\frac{\vec{P}_{\scr \bf R}^2}{2\mu_{\scr PT}} + V_{{\scr 1T}}\left( \vec{R}, \vec{r} \right) + V_{{\scr 2T}}\left( \vec{R},\vec{r} \right) + h_0,
\label{Hclass-1}
\end{equation}
where $h_0$ is the intrinsic Hamiltonian of the projectile,
\begin{equation}
h_0=\frac{\vec{p}_{\scr \bf r}^2}{2\mu_{\scr 12}} + V_{{\scr 12}}\left( \vec{r} \right).
\end{equation}
In the above equations $\vec{P}_{\scr R}$ and $\vec{p}_{\scr r}$ are the momenta associated with $\vec{R}$ and $\vec{r}$, and 
$\mu_{\scr PT}$ and $\mu_{\scr 12}$ are the reduced masses of the projectile-target system and of the two clusters of the
projectile. This Hamiltonian has both bound and scattering eigenstates, and its ground state is denoted by $\varphi_0$.
The infinite space of scattering states is approximated by a finite set of wave packets, referred to as {\it bins}, by the CDCC 
method~\cite{SYK83,SYK86,AIK87}.

The projectile-target motion is treated by classical mechanics. The trajectory, $\vec{r}(t)$, is obtained solving the classical 
equations of motion with the Hamiltonian 
\begin{equation}
\mathcal{H}_{\scr class} (\vec{R}) = \frac{\vec{P}_{\scr \bf R}^2}
{2\mu_{\scr PT}}  + \overline{V}(\vec{R}) ,
\label{Hcl}
\end{equation}
with
\begin{equation}
\overline{V}(\vec{R}) = \int d^3\vec{r}\ \big[  V_{{\scr 1T}}\left( \vec{R}, \vec{r} \right) + V_{{\scr 2T}}\left( \vec{R},\vec{r} \right) \big]\ 
\left| \varphi_0(\vec{r}) \right|^2.
\end{equation}
The intrinsic wave function of the projectile is expanded over the  bound and continuum-discretized eigenstates of $h_0$, and the
coefficients are found by solving the Schr\"odinger equation in the $\vec{r}$-space, with the time-dependent Hamiltonian 
\begin{eqnarray} 
h(t) &=& h_0 + V_{{\scr 1T}}\left( t, \vec{r} \right) + V_{{\scr 2T}}\left( t,\vec{r} \right) \nonumber\\
   &\equiv&  h_0 +
V_{\scr 1T}\left( \vec{R} (t), \vec{r} \right) + V_{\scr 2T}\left( \vec{R} (t),\vec{r} \right)\nonumber.
\end{eqnarray} 

The CF cross section is then found multiplying the final populations of the bound states by transmission coefficients of the whole projectile through the 
barrier of $\overline{V}(\vec{R})$. The ICF cross section for each fragment, depends on the population of continuum states and transmission coefficients 
of the fragment through its interaction potential with the target. The association of CF and ICF with bound states and continuum states of the projectile 
was proposed in the quantum mechanical calculations of Hagino {\it et al.}~\cite{HVD00}, which is discussed later in this section. The semiclassical 
method was used to calculate CF and ICF cross sections in collisions of $^{6,7}$Li with $^{209}$Bi \cite{MCD14}, $^{197}$Au and $^{159}$Tb 
\cite{KCD18}. The overall agreement between the theoretical CF cross sections and the data at near- and above barrier energies was reasonably good, 
while the predictions of ICF cross sections were poorer. \\

The most reliable calculations of fusion cross sections in collisions of weakly bound nuclei are based on the CDCC method. Standard CDCC calculations
give only $\sigma_{\scr TF}$~\cite{KKR01,DTB03,JPK14,DDC15}. It is extracted from  $\sigma_{\scr R}$ through the relation,
\begin{equation}
\sigma_{\scr TF} = \sigma_{\scr R} - \sum_{\gamma \ne 0} \sigma_{\gamma} ,
\label{sigTF-0}
\end{equation}
where $\gamma\ne 0$ correspond to the nonelastic channels (the elastic is labelled by $\gamma = 0$), both bound and unbound (continuum discretized).
In Eq.~(\ref{sigTF-0}), $\sigma_{\scr R} $ is the total reaction cross section, given by the expansion
\begin{equation}
\sigma_{\scr R} = \frac{\pi}{k^2}\  \sum_{J} (2J+1)\ \left[\, 1 -  \left| S_0(J)\right|^2\, \right],
\label{sigR-CDCC}
\end{equation}
where $S_0(J)$ is the elastic S-matrix in a collision with total angular momentum $J$.\\

However, Hagino {\it et al.}~\cite{HVD00} proposed a CDCC based method that gives individual CF and ICF cross section in collisions of weakly
bound nuclei. In their method, the TF cross section was evaluated directly by the expression,
\begin{equation}
\sigma_{\scr TF} = \frac{\pi}{k^2}\  \sum_{J} (2J+1)\ P_{\scr TF}(J).
\label{sigTF-Hag}
\end{equation}
The TF probabilities were expressed as a sum of contributions from all channels (bound and unbound) involved in the CDCC equations. In each channel
and for each $J$, this probability was determined by the flux that reaches the inner region of the barrier. The CF cross section was then associated with
the contributions from the elastic channel (and from inelastic channels for 
bound excited states, if any), whereas the ICF cross section was
given by the contribution from the bins. This method was used to study CF and ICF in the $^{11}$Be + $^{208}$Pb system. Their CDCC calculations 
were performed with real potentials, using the ingoing waves boundary conditions (IWBC) to account for fusion absorption. Comparing the obtained CF cross 
section with the fusion cross section without breakup couplings, they found enhancement at sub-barrier energies and suppression above the barrier.\\

Diaz-Torres and Thompson~\cite{DiT02} used the same approach to evaluate CF and ICF for the same system. However, 
instead of IWBC, they used a short-range 
imaginary potential, $W(R)$, depending exclusively on the distance between the centers of the projectile and the target. Thus, this potential is diagonal in 
channel space. The TF cross section was then given by the well known expression~\cite{CaH13},
\begin{equation} 
\sigma_{\scr TF} =   \frac{k}{E}\ \sum_{\gamma,\gamma^\prime} \
 \left\langle \psi_\gamma \left|\, - W_{\gamma \gamma^\prime}\,
 \right|   \psi_{\gamma^\prime} \right\rangle ,
\label{TF-1}
\end{equation}
where $W_{\gamma \gamma^\prime}$ are the matrix-elements of the imaginary potential and $ \psi_\gamma$ is the projectile-target relative wave
function in channel $\gamma$. Since $W_{\gamma\gamma^\prime} =  W_{\gamma \gamma}\cdot \delta_{\gamma,\gamma^\prime}$, the above expression reduces to
\begin{equation} 
\sigma_{\scr TF} =  \sum_{\gamma} \ \sigma_{\scr TF}^{\scr (\gamma)}
\label{TF-2}
\end{equation}
with
\begin{equation} 
\sigma_{\scr TF}^{\scr (\gamma)} =  \frac{k}{E}\  \left\langle \psi_\gamma \left|\,  W_{\gamma}\,
\right|  \psi_{\gamma} \right\rangle.
\label{TF-3}
\end{equation}
The CF and ICF components of $\sigma_{\scr TF}$ where then evaluated by the method of Ref.~\cite{HVD00}. That is, the cross sections
$\sigma_{\scr CF}$ and $\sigma_{\scr ICF}$ were obtained restricting the sum of Eq.~(\ref{TF-2}) to bound and to
unbound channels, respectively. The calculations of Ref.~\cite{DiT02} adopted a larger continuum space and took into account continuum-continuum
couplings, neglected in Ref.~\cite{HVD00}, but qualitatively, they lead to the same conclusion, namely: enhancement of CF at sub-barrier energies and 
suppression above the barrier. \\

The calculations of Ref.~\cite{HVD00,DiT02} have a limitation. They cannot be used in collisions of projectiles that break up into fragments of comparable masses. 
The association of ICF with unbound channels, where the two fragments tend to be far apart,  is based on the assumption that the center of mass of the 
heavier fragment is very close to the center of the projectile. In this way, the imaginary potential absorbs the heavier fragment but not the lighter one. The
method is justified in the case of $^{11}$Be, that breaks up into $^{10}$Be and a neutron, since the mass of the former 
is ten times larger than that of the latter.
However, it cannot be used in collisions of nuclei like $^7$Li, that breaks up into a triton and an alpha particle. In this case the mass ratio is 4/3. To deal with this kind 
of weakly bound nuclei, it is necessary to use individual imaginary potentials for the two fragments. This generalisation has been carried out in a recent paper 
by Rangel {\it et al.}~\cite{RCL20}. The extended method was applied to the $^7$Li + $^{209}$Bi system and the resulting CF and ICF cross sections were compared to the data of
Dasgupta {\it et al.}~\cite{DHH02,DGH04}. The theoretical predictions for both cross sections were in very good agreement with the data.\\

Another quantum mechanical method to evaluated $\sigma_{\scr CF}$ was 
recently proposed by Lei and Moro~\cite{LeM19}. In a collision of a weakly bound projectile 
composed of fragments $c_{\scr 1}$ and $c_{\scr 2}$, the CF cross section was extracted from the expression,
\begin{equation} 
\sigma_{\scr R} \simeq  \sigma_{\scr CF} + \sigma_{\rm inel} + \sigma_{\scr EBU} + \sigma_{\scr NEB}^{\scr (c_1)} + \sigma_{\scr NEB}^{\scr (c_2)},
\end{equation}
where $\sigma_{\scr R}$ is the total reaction cross section, $\sigma_{\rm inel}$ is the cross section for inelastic excitations, $\sigma_{\scr EBU}$ is the cross section 
for elastic breakup of the projectile, and $ \sigma_{\scr NEB}^{\scr (c_i)}$ ($i=1,2$), is the cross section for nonelastic breakup, where fragment $c_i$ emerges from 
the interaction region and the target does not remain in its ground state (the other fragment may be captured by the target or collide inelastically with it).
These cross sections were evaluated by different theoretical methods:  $\sigma_{\scr R}$ and $\sigma_{\scr EBU}$ were determined by CDCC calculations 
with appropriate imaginary potentials, $\sigma_{\rm inel}$ was obtained through a standard coupled channel calculation involving the main collective states, 
and the nonelastic breakup cross sections were calculated by the spectator-participant inclusive breakup model of Ichimura, Austern, and Vincent 
(IAV) \cite{AIK87,AuV81,IAV85}.  The IAV was used to calculate CF cross sections for the $^{6,7}$Li + $^{209}$Bi systems. The resulting cross sections at 
above-barrier energies were shown to be in good agreement with the data of Dasgupta {\it et al.}~\cite{DHH02,DGH04}. 
We should also mention the work of Parkar {\it et al.}~\cite{PJK16}, were CF and ICF cross sections for the $^{6,7}$Li + $^{209}{\rm Bi}, ^{198}{\rm Pt}$ systems
were obtained in approximate calculations with different short-range imaginary potentials. Their theoretical predictions were in reasonable agreement with the data 
of Refs.~\cite{DHH02,DGH04} and \cite{SNL09,SND13}, respectively.\\

There are still other promising theoretical methods which have not yet been developed to the point of making quantitative predictions of CF and ICF data. Hashimoto 
{\it et al.}~\cite{HOC09} proposed a CDCC-based method where the fusion cross sections were given by radial integrals of the fragment-target imaginary potentials, 
expressed in terms their separation vectors, $\vec{r}_1$ and $\vec{r}_2$. Then, the CF and ICF cross sections were respectively assigned to contributions from small 
and large values of $r_1$ and $r_2$ in the integrand. The same idea was used in a qualitative one-dimensional model proposed by Boseli and Diaz-Torres \cite{BoD14,BoD15}, using position projected operators. The same authors proposed a time-dependent wave-packet approach to described collisions of three body 
systems, and performed calculations of $\sigma_{\scr CF}$ and $\sigma_{\scr ICF}$ for the $^6$Li + $^{209}$Bi system, within a schematic one-dimensional model. 
Their time-dependent method has also been used to study collisions of tightly bound systems~\cite{DiW18,VoD19}.

\section{Elastic Heavy-Ion Scattering}

\subsection{Rainbow and glory scattering}

Let us next discuss elastic heavy-ion scattering. In the classical mechanics, 
the differential cross section is given by 
\begin{equation}
\frac{d\sigma_{\rm cl}}{d\Omega}=\frac{\lambda}{k^2\sin\theta}
\left(\frac{d\Theta(\lambda)}{d\lambda}\right)^{-1},
\label{sigma-cl}
\end{equation}
where $\theta$ is a scattering angle, $k$ is the wave number, and $\lambda$ is the 
angular momentum. $\Theta(\lambda)$ is the scattering angle as a function of $\lambda$, 
that is, the deflection function. 
The classical cross section, Eq. (\ref{sigma-cl}), diverges at 
$d\Theta(\lambda)/d\lambda=0$ as well as at $\sin\theta=0$, which are referred to as 
{\it rainbow} scattering and {\it glory} scattering, respectively. 
These are caustics in a sense that many angular momenta, $\lambda$, contribute 
coherently to scattering for 
a particular scattering angle $\theta$ and its vicinity. 

In the semi-classical approximation, the deflection function $\Theta(\lambda)$ is 
related to scattering phase shifts, $\delta_l$, for a partial wave $l$ as 
\cite{Bri85}, 
\begin{equation}
\Theta(\lambda)=2\frac{d\delta_l}{dl}\sim 2(\delta_{l+1}-\delta_l)
\end{equation}
with $\lambda=l+1/2$. 
Here, the phase shift $\delta_l$ is a sum of a nuclear phase shift $\delta^{(N)}_l$ 
and the Coulomb phase shift $\delta^{(C)}_l$, that is, 
$\delta_l=\delta^{(N)}_l+\delta^{(C)}_l$. 

\begin{figure}[tb]
\begin{center}
\includegraphics[clip,scale=0.5]{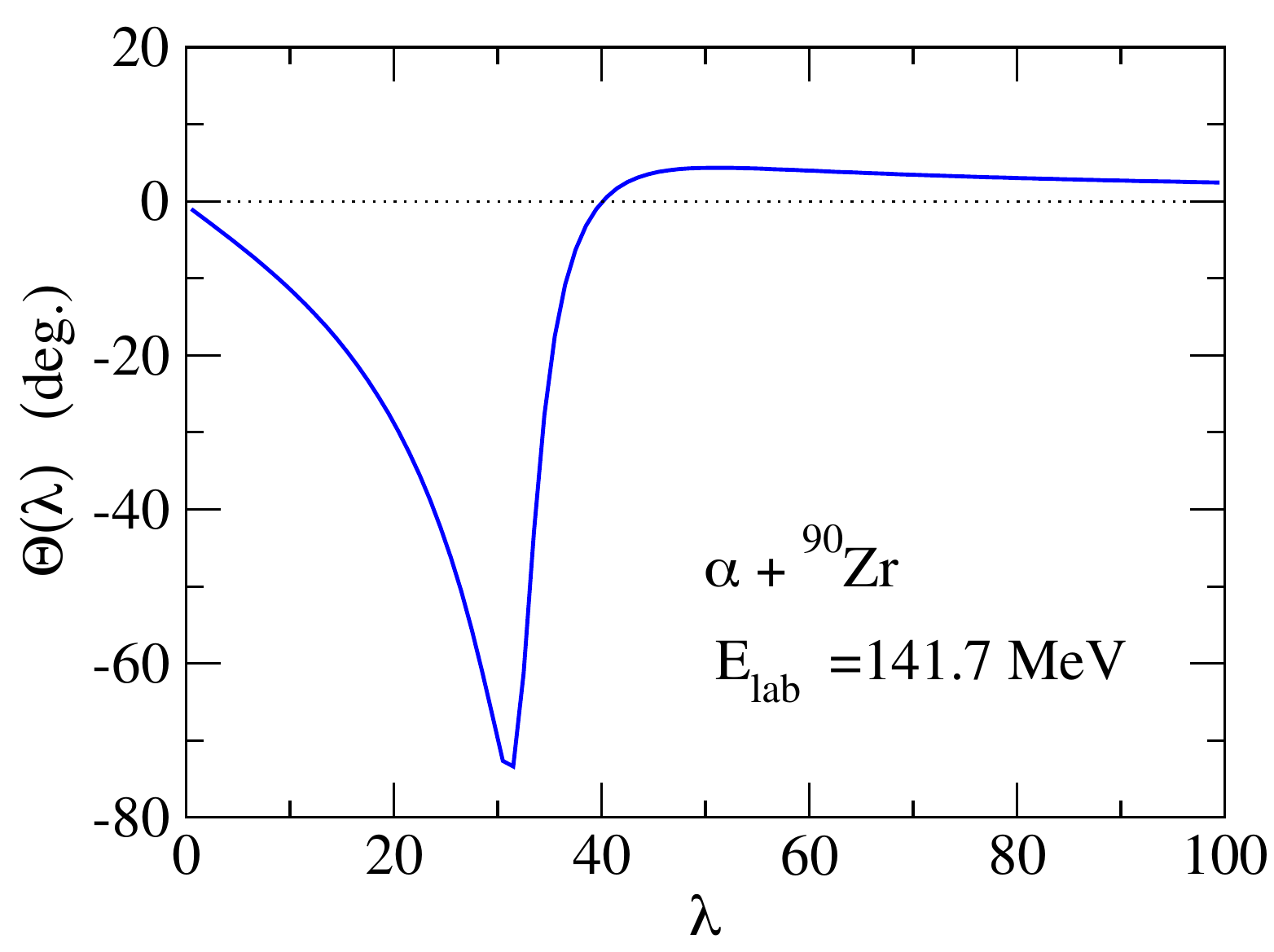}
\caption{The deflection function for the $\alpha$+$^{90}$Zr scattering 
at energy $E_{\rm lab}=141.7$ MeV. 
It is obtained by neglecting the imaginary part of an optical potential 
of the Woods-Saxon form. 
}
\label{fig:deflection}       
\end{center}
\end{figure}

Fig. \ref{fig:deflection} shows a deflection function 
for $\alpha$+$^{90}$Zr scattering at 
$E_{\rm lab}=141.7$ MeV in the laboratory frame. 
The phase shifts are evaluated with a Woods-Saxon potential with 
the depth, the range, and the diffuseness parameters of $V_0=117.5$ MeV, 
$R=1.267\times 90^{1/3}$ fm, and $a=0.783$ fm, respectively \cite{GSB74}. 
No imaginary part is included in the potential to draw the deflection function. 
One can see that the rainbow scattering takes place, at $\lambda=50.5$ with 
a rainbow angle of $\Theta_R=4.34$ degrees 
and 
at $\lambda= 31.5$ 
with $\Theta_R=-73.4$ degrees. 
The former is due to a balance between the repulsive Coulomb 
interaction and an attractive nuclear interaction, and is referred to as 
the Coulomb rainbow scattering. On the other hand, the latter is due to purely 
a nuclear interaction, and is referred to as nuclear rainbow scattering. 
Scattering angles larger than $|\Theta_R|$ are forbidden classically, but they 
are allowed in quantum mechanics due to the diffraction of a wave function; this 
is called shadow scattering. 
In addition to the rainbows, the deflection function crosses 
zero at $\lambda\sim 40$, that is 
the condition for (forward) glory scattering. 

\begin{figure}[tb]
\begin{center}
\includegraphics[clip,scale=0.6]{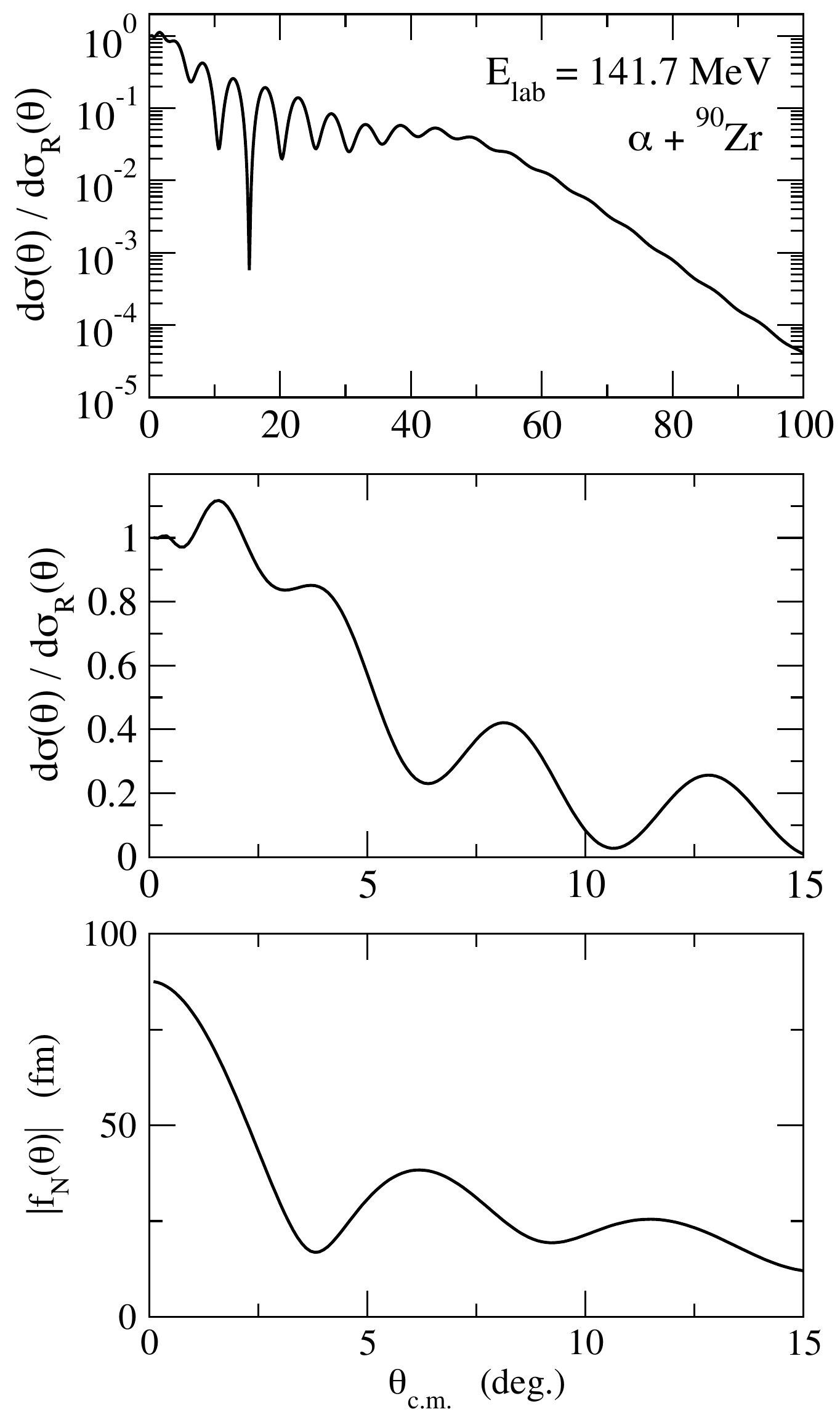}
\caption{(Top) The angular distribution for elastic 
$\alpha$+$^{90}$Zr 
scattering at $E_{\rm lab}=141.7$ MeV. 
It is given as a ratio to the Rutherford cross sections, $d\sigma_R(\theta)$. 
(Middle) The same as the top panel, but plotted in the linear scale for 
forward scattering angles. 
(Bottom) The absolute value of the nuclear scattering amplitude. 
}
\label{fig:4he90zr}       
\end{center}
\end{figure}

The top panel of Fig. \ref{fig:4he90zr} shows the ratio of 
the differential cross sections to the Rutherford cross sections. 
To compute the cross sections, we include the imaginary part of the 
potential, with a Woods-Saxon parameterization with $W=21.02$ MeV, 
$R_w=1.267\times 90^{1/3}$ fm, and $a_w=0.783$ fm, together with 
the charge radius of 
$R_w=1.3\times 90^{1/3}$ fm for the Coulomb interaction \cite{GSB74}. 
This potential well reproduces the experimental data at $E_{\rm lab}=141.7$ MeV 
 \cite{GSB74}. 
The cross sections show a bump around 
$\theta_{\rm c.m.}=45$ degrees, with an exponential fall off at larger scattering 
angles. This is a clear manifestation of the nuclear rainbow scattering. 
At angles smaller than the rainbow angle, there are two angular momenta which 
lead to the same scattering angle, as can bee seen in the deflection function 
shown in Fig. \ref{fig:deflection}. These two components interfere with each other, 
leading to a characteristic interference pattern given by the Airy 
function \cite{Bri85}. 

The middle panel of Fig. \ref{fig:4he90zr} shows the differential 
cross sections at forward angles in the linear scale. 
Here also one can see a characteristic interference pattern around 
$\theta_{\rm c.m.}=1.6$ degrees, which can be interpreted as the Coulomb 
rainbow scattering. 

The bottom panel of Fig. \ref{fig:4he90zr} shows the absolute value of 
the nuclear scattering amplitude, $f_N(\theta)$. 
This quantity is enhanced at $\theta_{\rm c.m.}=0$ degrees, with characteristic 
oscillations. This is a manifestation of the glory scattering, for which the 
interference is originated from the contribution of the near-side component 
with $\theta$ and that of the far-side component with $-\theta$. 
In the semi-classical approximation, the nuclear scattering amplitude at forward 
angles is given by \cite{Bri85,CaH13}, 
\begin{equation}
f_N(\theta)\propto \sqrt{\frac{2\pi\theta}{\sin\theta}}\,J_0(\lambda_g\theta), 
\label{glory}
\end{equation}
where $\lambda_g$ is the glory angular momentum at which the deflection function 
crosses zero. 
That is, the glory scattering is characterized by the zero-th order Bessel function, 
$J_0$. 

Incidentally, 
the Rutherford cross section diverges at $\theta=0$ and it may not be 
straightforward to 
probe the forward glory scattering experimentally. Yet, one can still use the 
generalized optical theorem, that is, the sum-of-differences (SOD) method, to extract 
the nuclear scattering 
amplitude \cite{UPH98,UPH99,HNV82,HuS84a}. 
The SOD cross section is defined as 
\begin{equation}
\sigma_{\rm SOD}(\theta)=2\pi\int^\pi_\theta
\left(\frac{d\sigma_R}{d\Omega'}-\frac{d\sigma}{d\Omega'}\right)\,
\sin\theta'd\theta', 
\label{sod}
\end{equation} 
and is related to the total reaction cross section $\sigma_R$ and the nuclear 
scattering amplitude as 
\begin{eqnarray}
&&\sigma_{\rm SOD}(\theta)\sim\sigma_{R}  \nonumber \\
&&-\frac{4\pi}{k}|f_N(\theta)|\sin\left({\rm arg} f_N(\theta)-2\delta^{(C)}_0
+2\eta\ln\sin\frac{\theta}{2}\right). 
\end{eqnarray}
where $\eta$ is the Sommerfeld 
parameter. Here we have neglected small correction terms, which can be ignored 
when $\theta$ is small. 
Notice that one can also access to fusion cross sections by taking 
\begin{equation}
\sigma'_{\rm SOD}(\theta)=2\pi\int^\pi_\theta
\left(\frac{d\sigma_R}{d\Omega'}-\frac{d\sigma_{\rm qel}}{d\Omega'}\right)\,
\sin\theta'd\theta' 
\end{equation} 
in Eq. (\ref{sod}), where $d\sigma_{qel}/d\Omega$ is the quasi-elastic cross section 
defined as a sum of elastic, inelastic, transfer, and breakup cross sections. 
In this case, $\sigma'_{\rm SOD}$ is related to fusion cross sections, $\sigma_{\rm fus}$, 
as \cite{HaR15}, 
\begin{eqnarray}
&&\sigma'_{\rm SOD}(\theta)\sim\sigma_{\rm fus}  \nonumber \\
&&-\frac{4\pi}{k}|f_N(\theta)|\sin\left({\rm arg} f_N(\theta)-2\sigma_0
+2\eta\ln\sin\frac{\theta}{2}\right). 
\end{eqnarray}

\begin{figure}[tb]
\begin{center}
\includegraphics[clip,scale=0.6]{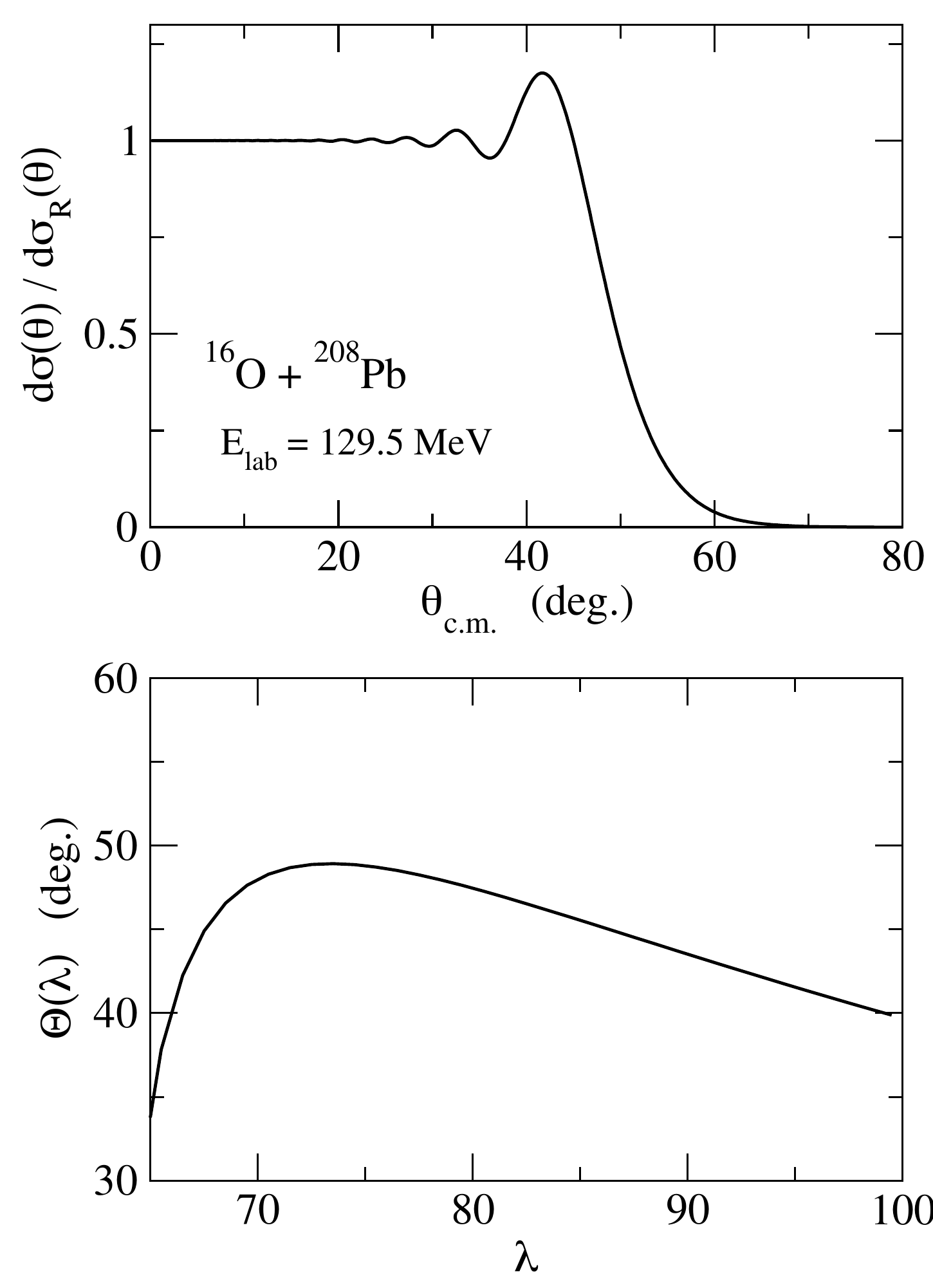}
\caption{(Upper panel) The angular distribution for elastic 
$^{16}$O+$^{208}$Pb  
scattering at $E_{\rm lab}=129.5$ MeV. 
It is given as a ratio to the Rutherford cross sections, $d\sigma_R(\theta)$. 
(Lower panel) The corresponding deflection function as a function of the 
angular momentum $\lambda$. 
}
\label{fig:16o208pb}       
\end{center}
\end{figure}

In the $\alpha$+$^{90}$Zr scattering shown in Fig. \ref{fig:deflection}, the 
Coulomb rainbow angle is small and the forward scattering is actually 
affected both by the Coulomb rainbow scattering and by the glory scattering. 
The Coulomb rainbow is more clearly seen when a heavier nucleus 
is used as a projectile. In order to demonstrate this, 
the upper panel of Fig. \ref{fig:16o208pb} 
shows the elastic cross sections for the $^{16}$O+$^{208}$Pb system at 
$E_{\rm lab}=129.5$ MeV. The deflection function is also shown in 
the lower panel. To this end, we use the optical potential given in 
Ref. \cite{BFG75} (the deflection function is obtained using only the 
real part while the cross sections are calculated including both the 
real and the imaginary parts). The deflection function has a maximum 
at $\lambda=73.5$ with the rainbow angle of 48.9 degrees. 
The scattering cross sections exhibit a characteristic Fresnel oscillation pattern, 
which can be interpreted as the Coulomb rainbow scattering 
(see Sec. 5.6 in Ref. \cite{Bri85} for a discussion on a relation 
between the Coulomb rainbow scattering and the Fresnel diffraction). 

\subsection{Glory in the shadow of rainbow}

\begin{figure}[tb]
\begin{center}
\includegraphics[clip,scale=0.3]{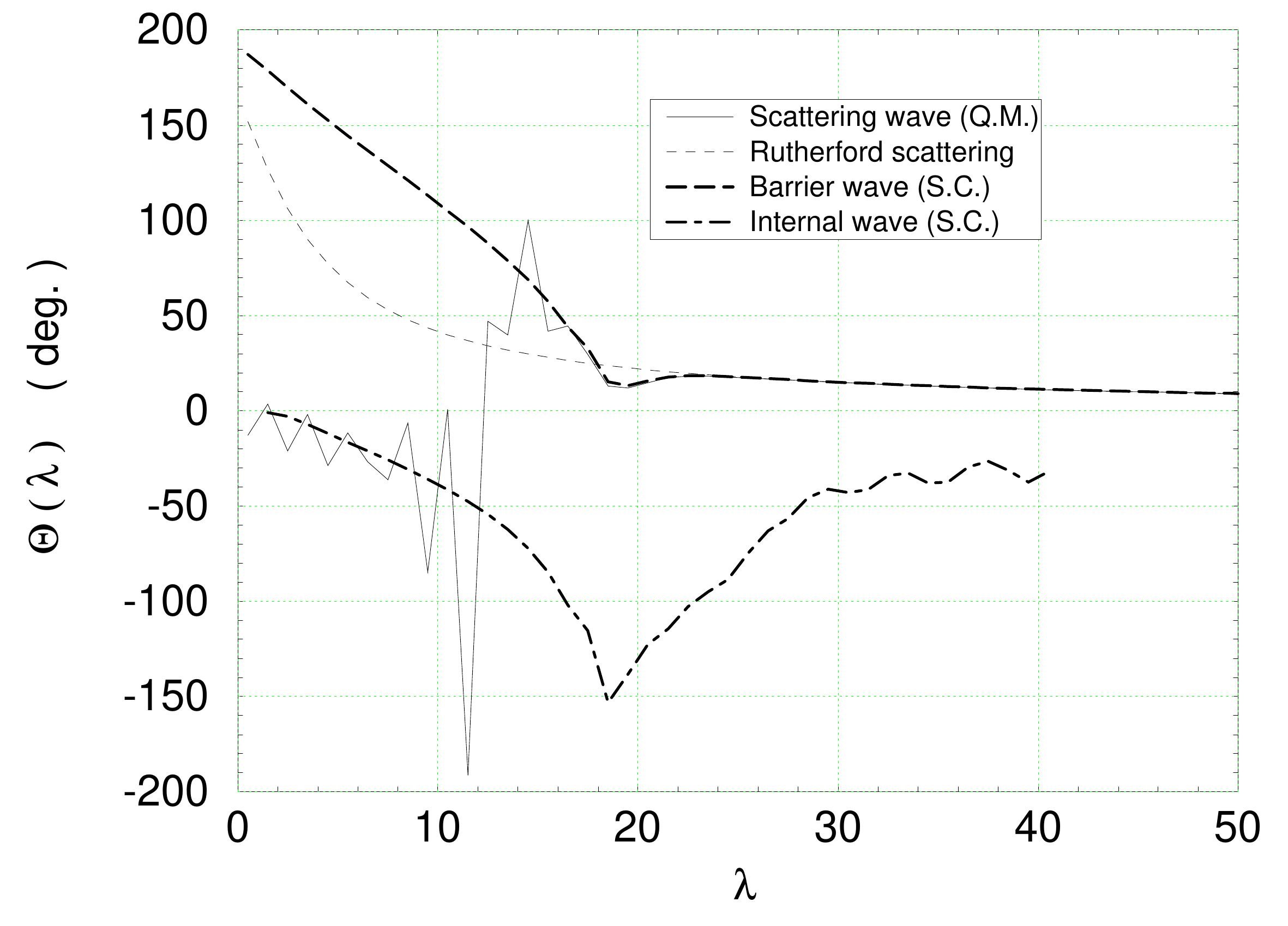}
\includegraphics[clip,scale=0.3]{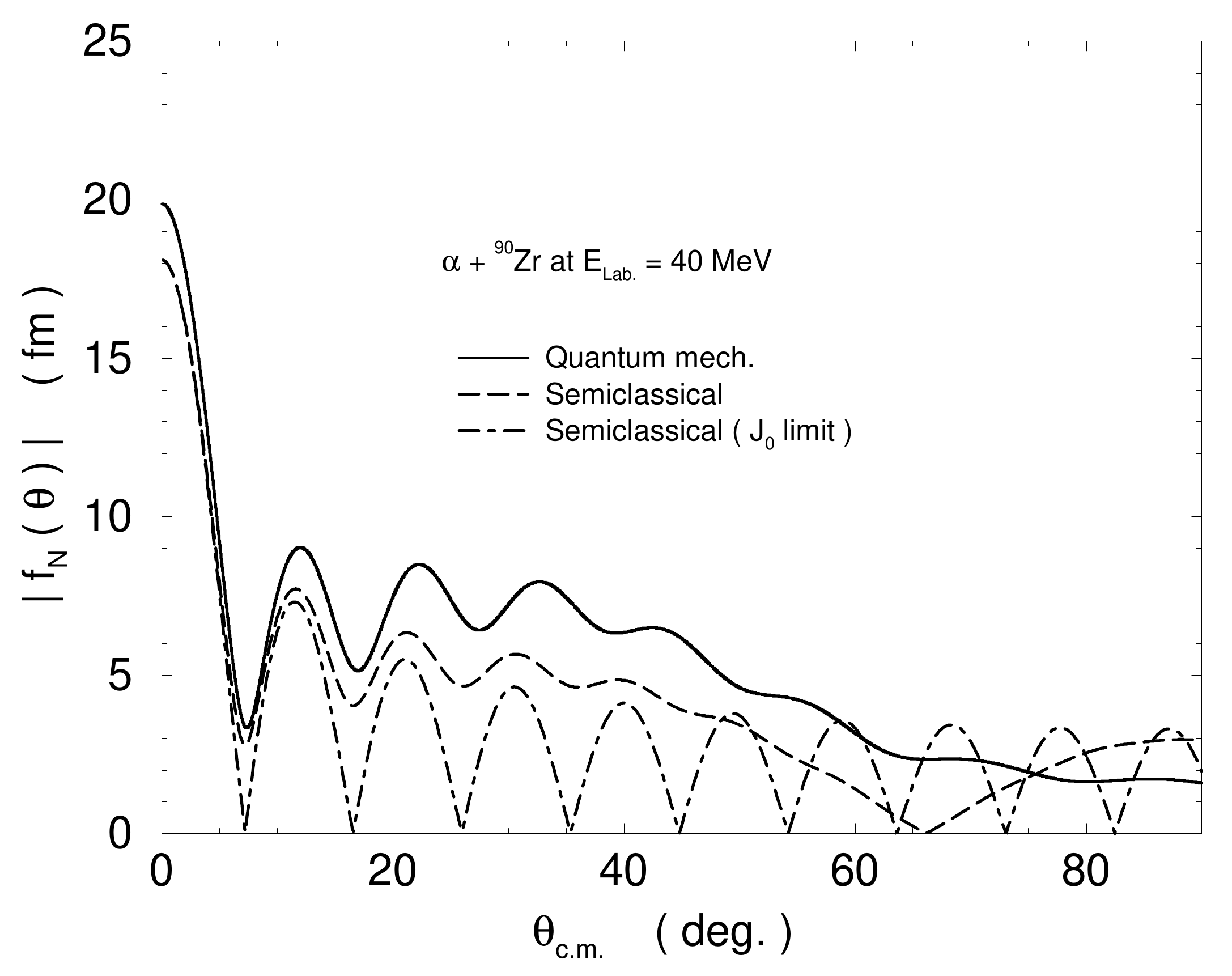}
\caption{(Upper panel) The 
deflection function for $\alpha+^{90}$Zr scattering at $E_{\rm lab}=40$ MeV. 
The solid line is obtained with a quantum mechanical calculation. 
The dashed 
and the dot-dashed lines denote the deflection function for the barrier wave 
and the internal wave, respectively, 
obtained with the semi-classical approximation. 
The thin dashed line shows the deflection 
function for the pure Coulomb scattering. 
(Lower panel) The absolute value of the nuclear scattering amplitude. 
The solid line shows the result of the quantum mechanical calculation, while 
the dashed line is obtained with the semi-classical approximation. The dot-dashed 
line is obtained with a semi-classical formula which is valid at very 
small scattering angles. 
Taken from Ref. \cite{UPH99}. 
}
\label{fig:glory}       
\end{center}
\end{figure}

Even though Fig. \ref{fig:4he90zr} clearly shows the features of the nuclear 
rainbow and the glory scattering, 
the interference pattern becomes much more 
complex as the incident energy decreases. 
Moreover, the effect of absorption 
becomes more important. In particular, the deflection function may not cross 
zero but a minimum appears with a positive rainbow angle. 
This is illustrated in the upper panel of Fig. \ref{fig:glory} 
for $\alpha+^{90}$Zr scattering at $E_{\rm lab}=40$ MeV.  
In this figure, the deflection function is decomposed into the barrier wave 
contribution and the internal wave contribution, where the former 
corresponds to the flux reflected at the barrier while the latter 
corresponds to the flux reflected at the innermost turning point \cite{BrT77a}. 
The 
deflection function evaluated with a quantal calculation (the solid line) 
indicates that 
there is a crossover from the internal wave (the dot-dashed line) 
to the barrier wave (the dashed line) as the angular 
momentum increases, and that 
the barrier wave is responsible for the nuclear rainbow scattering 
for this system. An important feature is that the effect of 
nuclear interaction is not strong enough for the barrier wave 
so that the deflection function does not cross zero before it bends when the 
angular moment decreases from the angular momentum for the Coulomb rainbow 
scattering. Interestingly, 
as shown in the lower panel of Fig. \ref{fig:glory}, 
the nuclear scattering amplitude 
still shows an enhancement at $\theta=0$ as in the glory scattering shown in 
the bottom panel of Fig. \ref{fig:4he90zr}. 
That is, 
the nuclear scattering amplitude exhibits a similar behavior to glory scattering 
even thought the deflection function does not cross zero. 

In order to interpret this phenomenon, one of us (M.U.), together with 
Mahir Hussein, introduced a novel concept of glory in the shadow of 
rainbow \cite{UPH98,UPH99}. 
That is, when the deflection function does not cross zero, the zero scattering 
angle corresponds to the shadow region of the nuclear rainbow scattering. 
An important point is that 
the nuclear rainbow 
can still affect the zero angle scattering if the rainbow angle is small 
because of the diffractive nature of the wave function. 
Using the semi-classical approximation, one can actually 
derive the expression for the 
nuclear scattering amplitude for glory in the shadow of rainbow as \cite{UPH98,UPH99}, 
\begin{equation}
f_N(\theta)\propto \sqrt{\frac{2\pi\theta}{\sin\theta}}\,
(A_+(\theta)J_0(\lambda_{\rm NR}\theta)
+iA_-(\theta)J_0(\lambda_{\rm NR}\theta)) 
\label{glory-rainbow}
\end{equation}
with 
\begin{eqnarray}
A_\pm(\theta)&=&{\rm Ai}(\xi_1)\pm {\rm Ai}(\xi_2),  \\
\xi_1&=& \left(\frac{\Theta''_{\rm NR}}{2}\right)^{-1/3}(\theta_{\rm NR}+\theta), \\
\xi_2&=& \left(\frac{\Theta''_{\rm NR}}{2}\right)^{-1/3}(\theta_{\rm NR}-\theta), 
\end{eqnarray}
where ${\rm Ai}$ denotes the Airy function and 
$\lambda_{\rm NR}$ is the nuclear rainbow angular momentum at which the deflection 
function takes a minimum. 
$\theta_{\rm NR}$ and 
$\Theta''_{\rm NR}$ are the scattering angle and the curvature of the 
deflection function at $\lambda=\lambda_{\rm NR}$, respectively. 
Notice that 
Eq. (\ref{glory-rainbow}) is indeed similar to the formula for glory scattering, 
Eq. (\ref{glory}), with a replacement of the glory angular momentum $\lambda_g$ with 
the nuclear rainbow angular momentum $\lambda_{\rm NR}$. 
In the bottom panel of Fig. \ref{fig:glory}, one can see that the semi-classical 
formula (the dashed line) well reproduces the quantum mechanical calculation (the 
solid line), supporting the concept of glory in the shadow of rainbow. 

\section{Summary} 

We have discussed the semi-classical approaches to low-energy 
heavy-ion reactions, focusing on heavy-ion fusion reactions of neutron-rich 
nuclei and the phenomena of nuclear rainbow and glory scattering in 
elastic heavy-ion scattering. Mahir Hussein significantly 
contributed to both topics, enhancing our understanding of the 
nature of low-energy heavy-ion reactions.  
We mention that fusion of neutron-rich nuclei are still important in connection to 
fusion in neutron stars, as well as syntheses of superheavy nuclei, 
especially attempts to reach the island 
of stability. Towards these goals, 
there are still many interesting topics to clarify in fusion of 
neutron-rich nuclei, such as an interplay between breakup and transfer. 

\begin{figure}[tb]
\begin{center}
\includegraphics[clip,scale=0.4,angle=-1]{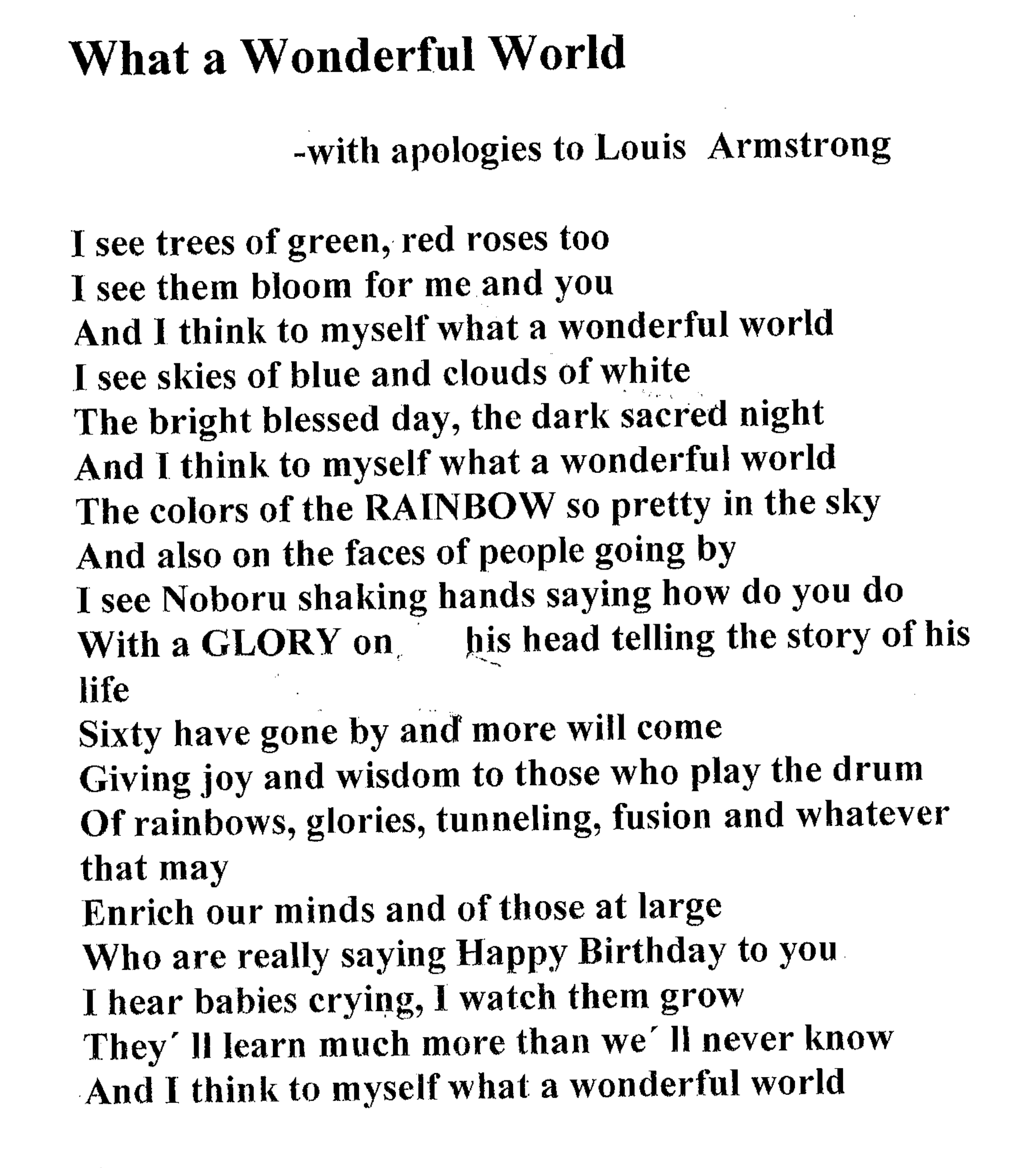}
\caption{
A copy of a transparency which Mahir Hussein showed at the 
workshop on occasion of Noboru Takigawa's 60th birthday (November 2003, 
Sendai, Japan). 
}
\label{fig:hussein}       
\end{center}
\end{figure}

We would like to close this paper by showing 
a parody of ``what a wonderful world'' which Hussein showed at the meeting 
for Takigawa (see Fig. \ref{fig:hussein} with slight modifications by us), since 
we feel that lyrics nicely reflects Hussein's nature as a nuclear physicist. 

\medskip

\begin{flushleft}
{\bf ``What a Wonderful World'' (with apologies to Louis Armstrong)} \\
\medskip

I see trees of green, red roses too \\
I see them bloom for me and you \\
And I think to myself what a wonderful world \\
I see skies of blue and clouds of white \\
The bright blessed day, the dark sacred night \\
And I think to myself what a wonderful world \\
The colors of the {\bf RAINBOW} so pretty in the sky \\
And also on the faces of people going by \\
I see Mahir shaking hands saying how do you do \\
With a {\bf GLORY} on his head telling the story of his life \\
I hear babies crying, I watch them grow \\
They'll learn much more than we'll never know \\
And I think to myself what a wonderful world. 

\end{flushleft}

\bibliographystyle{epj}
\bibliography{fusbreak-2020_vs13} 

\begin{thebibliography}{75}

\bibitem{Bri85}
D.M. Brink, \emph{Semi-classical Methods for Nucleus-Nucleus Scattering}
  (Cambridge University Press, Cambridge, England, 1985)

\bibitem{CaH13}
L.F. Canto, M.S. Hussein, \emph{Scattering Theory of Molecules, Atoms and
  Nuclei} (World Scientific Publishing Co. Pte. Ltd., Singapore, 2013)

\bibitem{FrL96}
P.~Fr\"obrich, R.~Lipperheide, \emph{Theory of nuclear reactions} (Clarendon
  Press, Oxford, 1996)

\bibitem{BrW04}
R.A. Broglia, A.~Winther, \emph{Heavy Ion Reactions} (Westview Press,
  Cambrigde, MA, 2004)

\bibitem{CGD06}
L.F. Canto, P.R.S. Gomes, R.~Donangelo, M.S. Hussein, Phys. Rep. \textbf{424},
  1 (2006)

\bibitem{CGD15}
L.F. Canto, P.R.S. Gomes, R.~Donangelo, J.~Lubian, M.S. Hussein, Phys. Rep.
  \textbf{596}, 1 (2015)

\bibitem{HPC92}
M.S. Hussein, M.P. Pato, L.F. Canto, R.~Donangelo, Phys. Rev. C \textbf{46},
  377 (1992)

\bibitem{HuM84}
M.S. Hussein, K.W. McVoy, Rep. Prog. Part. Nucl. Phys. \textbf{12}, 103 (1984)

\bibitem{BrT77a}
D.M. Brink, N.~Takigawa, Nucl. Phys. A \textbf{279}, 159 (1977)

\bibitem{KOB07}
D.T. Khoa, W.~von Ortzen, H.G. Bohlen, S.~Ohkubo, J. Phys. G: Nucl. Part. Phys.
  \textbf{34}, R111 (2007)

\bibitem{UPH98}
M.~Ueda, M.P. Pato, M.S. Hussein, N.~Takigawa, Phys. Rev. Lett. \textbf{81},
  1809 (1998)

\bibitem{UPH99}
M.~Ueda, M.P. Pato, M.S. Hussein, N.~Takigawa, Nucl. Phys. A \textbf{648}, 229
  (1999)

\bibitem{THH85}
I.~Tanihata, H.~Hamagaki, O.~Hashimoto, Y.~Shida, N.~Yoshikawa, K.~Sugimoto,
  O.~Yamakawa, T.~Kobayashi, N.~Takahashi, Phys. Rev. Lett. \textbf{55}, 2676
  (1985)

\bibitem{BAG08}
C.~Bachelet, G.~Audi, C.~Gaulard, C.~Gu\'enaut, F.~Herfurth, D.~Lunney,
  M.~de~Saint~Simon, C.~Thibault, Phys. Rev. Lett. \textbf{100}, 182501 (2008)

\bibitem{NVS06}
T.~Nakamura, A.M. Vinodkumar, T.~Sugimoto, N.~Aoi, H.~Baba, D.~Bazin,
  N.~Fukuda, T.~Gomi, H.~Hasegawa, N.~Imai et~al., Phys. Rev. Lett.
  \textbf{96}, 252502 (2006)

\bibitem{BBH91}
C.A. Bertulani, G.~Baur, M.S. Hussein, Nucl. Phys. A \textbf{526}, 751 (1991)

\bibitem{BeE91}
G.F. Bertsch, H.~Esbensen, Annals of Physics \textbf{209}, 327 (1991)

\bibitem{HaS05}
K.~Hagino, H.~Sagawa, Phys. Rev. C \textbf{72}, 044321 (2005)

\bibitem{SaH15}
H.~Sagawa, K.~Hagino, Eur. Phys. J. A \textbf{51}, 102 (2015)

\bibitem{BeH07}
C.A. Bertulani, M.~S.~Hussein, Phys. Rev. C \textbf{76}, 051602 (2007)

\bibitem{HaS07}
K.~Hagino, H.~Sagawa, Phys. Rev. C \textbf{76}, 047302 (2007)

\bibitem{MMS05}
M.~Matsuo, K.~Mizuyama, Y.~Serizawa, Phys. Rev. C \textbf{71}, 064326 (2005)

\bibitem{Mat06}
M.~Matsuo, Phys. Rev. C \textbf{73}, 044309 (2006)

\bibitem{BaT98}
A.B. Balantekin, N.~Takigawa, Rev. Mod. Phys. \textbf{70}, 77 (1998)

\bibitem{DHR98}
M.~Dasgupta, D.~Hinde, N.~Rowley, A.~Stefanini, Ann. Rev. of Nucl. Part. Sci.
  \textbf{48}, 401 (1998)

\bibitem{HaT12}
K.~Hagino, N.~Takigawa, Prog. Theor. Phys. \textbf{128}, 1061 (2012)

\bibitem{BEJ14}
B.B. Back, H.~Esbensen, C.L. Jiang, K.E. Rehm, Rev. Mod. Phys. \textbf{86}, 317
  (2014)

\bibitem{MoS17}
G.~Montagnoli, A.M. Stefanini, Eur. Phys. J. A \textbf{53}, 169 (2017)

\bibitem{TaS91}
N.~Takigawa, H.~Sagawa, Phys. Lett. \textbf{B265}, 23 (1991)

\bibitem{SGT95}
H.~Sagawa, N.~Van~Giai, N.~Takigawa, M.~Ishihara, K.~Yasaki, Z. Phys. A
  \textbf{351}, 385 (1995)

\bibitem{DaV94}
C.H. Dasso, A.~Vitturi, Phys. Rev. C \textbf{50}, R12 (1994)

\bibitem{HVD00}
K.~Hagino, A.~Vitturi, C.H. Dasso, S.M. Lenzi, Phys. Rev. C \textbf{61}, 037602
  (2000)

\bibitem{CCS18}
K.S. Choi, M.K. Cheoun, W.~So, K.~Hagino, K.~Kim, Phys. Letters B \textbf{780},
  455 (2018)

\bibitem{IYN06a}
M.~Ito, K.~Yabana, T.~Nakatsukasa, M.~Ueda, Physics Letters B \textbf{637}, 53
  (2006)

\bibitem{LiR84}
R.~Lindsay, N.~Rowley, J. Phys. G: Nucl. Part. Phys. \textbf{10}, 805 (1984)

\bibitem{DLW83}
C.H. Dasso, S.~Landowne, A.~Winther, Nucl. Phys. \textbf{A405}, 381 (1983)

\bibitem{TKS93}
N.~Takigawa, M.~Kuratani, H.~Sagawa, Phys. Rev. C \textbf{47}, R2470 (1993)

\bibitem{HPC93}
M.S. Hussein, M.P. Pato, L.F. Canto, R.~Donangelo, Phys. Rev. \textbf{C47},
  2398 (1993)

\bibitem{SYK83}
Y.~Sakuragi, M.~Yahiro, M.~Kamimura, Prog. Theor. Phys. \textbf{70}, 1047
  (1983)

\bibitem{SYK86}
Y.~Sakuragi, M.~Yahiro, M.~Kamimura, Prog. Theoret. Phys. Suppl. \textbf{89},
  136 (1986)

\bibitem{AIK87}
N.~Austern, Y.~Iseri, M.~Kamimura, M.~Kawai, G.~Rawitscher, M.~Yashiro, Phys.
  Rep. \textbf{154}, 125 (1987)

\bibitem{DHH02}
M.~Dasgupta, D.J. Hinde, K.~Hagino, S.B. Moraes, P.R.S. Gomes, R.M. Anjos, R.D.
  Butt, A.C. Berriman, N.~Carlin, C.R. Morton et~al., Phys. Rev. C \textbf{66},
  041602(R) (2002)

\bibitem{HDH04}
K.~Hagino, M.~Dasgupta, D.J. Hinde, Nucl. Phys. \textbf{A738}, 475 (2004)

\bibitem{DHT07}
A.~Diaz-Torres, D.J. Hinde, J.A. Tostevin, M.~Dasgupta, L.R. Gasques, Phys.
  Rev. Lett. \textbf{98}, 152701 (2007)

\bibitem{Dia10}
A.~Diaz-Torres, J. Phys. G: Nucl. Part. Phys. \textbf{37}, 075109 (2010)

\bibitem{HDF02}
D.J. Hinde, M.~Dasgupta, B.R. Fulton, C.R. Morton, R.J. Wooliscroft, A.C.
  Berriman, K.~Hagino, Phys. Rev. Lett. \textbf{89}, 272701 (2002)

\bibitem{Dia11}
A.~Diaz-Torres, Comput. Phys. Commun. \textbf{182}, 1100 (2011)

\bibitem{DiQ18}
A.~Diaz-Torres, D.~Quraishi, Phys. Rev. C \textbf{97}, 024611 (2018)

\bibitem{DHB99}
M.~Dasgupta, D.J. Hinde, R.D. Butt, R.M. Anjos, A.C. Berriman, N.~Carlin,
  P.R.S. Gomes, C.R. Morton, J.O. Newton, A.~Szanto~de Toledo et~al., Phys.
  Rev. Lett. \textbf{82}, 1395 (1999)

\bibitem{DGH04}
M.~Dasgupta, P.R.S. Gomes, D.J. Hinde, S.B. Moraes, R.M. Anjos, A.C. Berriman,
  R.D. Butt, N.~Carlin, J.~Lubian, C.R. Morton et~al., Phys. Rev. C
  \textbf{70}, 024606 (2004)

\bibitem{AlW75}
K.~Alder, A.~Winther, \emph{Electromagnetic Excitations} (North-Holland,
  Amsterdam, 1975)

\bibitem{MCD14}
H.D. Marta, L.F. Canto, R.~Donangelo, Phys. Rev. C \textbf{89}, 034625 (2014)

\bibitem{KCD18}
G.D. Kolinger, L.F. Canto, R.~Donangelo, S.R. Souza, Phys. Rev. C \textbf{98},
  044604 (2018)

\bibitem{KKR01}
N.~Keeley, K.W. Kemper, K.~Rusek, Phys. Rev. C \textbf{65}, 014601 (2001)

\bibitem{DTB03}
A.~Diaz-Torres, I.J. Thompson, C.~Beck, Phys. Rev. C \textbf{68}, 044607 (2003)

\bibitem{JPK14}
V.~Jha, V.V. Parkar, S.~Kailas, Phys. Rev. C \textbf{89}, 034605 (2014)

\bibitem{DDC15}
P.~Descouvemont, T.~Druet, L.F. Canto, M.S. Hussein, Phys. Rev. C \textbf{91},
  024606 (2015)

\bibitem{DiT02}
A.~Diaz-Torres, I.J. Thompson, Phys. Rev. C \textbf{65}, 024606 (2002)

\bibitem{RCL20}
J.~Rangel, M.~Cortes, J.~Lubian, L.F. Canto, Phys. Lett. B \textbf{803}, 135337
  (2020)

\bibitem{LeM19}
J.~Lei, A.M. Moro, Phys. Rev. Lett. \textbf{122}, 042503 (2019)

\bibitem{AuV81}
N.~Austern, C.M. Vincent, Phys. Rev. C \textbf{23}, 1847 (1981)

\bibitem{IAV85}
M.~Ichimura, N.~Austern, C.M. Vincent, Phys. Rev. C \textbf{32}, 431 (1985)

\bibitem{PJK16}
V.V. Parkar, V.~Jha, S.~Kailas, Phys. Rev. C \textbf{94}, 024609 (2016)

\bibitem{SNL09}
A.~Shrivastava, A.~Navin, A.~Lemasson, K.~Ramachandran, V.~Nanal, M.~Rejmund,
  K.~Hagino, T.~Ishikawa, S.~Bhattacharyya, A.~Chatterjee et~al., Phys. Rev.
  Lett. \textbf{103}, 232702 (2009)

\bibitem{SND13}
A.~Shrivastava, A.~Navin, A.~Diaz-Torres, V.~Nanal, K.~Ramachandran,
  M.~Rejmund, S.~Bhattacharyya, A.~Chatterjee, S.~Kailas, A.~Lemasson et~al.,
  Phys. Lett. B \textbf{718}, 931 (2013)

\bibitem{HOC09}
S.~Hashimoto, K.~Ogata, S.~Chiba, M.~Yahiro, Prog. Theor. Phys. \textbf{122},
  1291 (2009)

\bibitem{BoD14}
M.~Boselli, A.~Diaz-Torres, J. Phys. G: Nucl. Part. Phys. \textbf{41}, 094001
  (2014)

\bibitem{BoD15}
M.~Boselli, A.~Diaz-Torres, Phys. Rev. C \textbf{92}, 044610 (2015)

\bibitem{DiW18}
A.~Diaz-Torres, M.~Wiescher, Phys. Rev. C \textbf{97}, 055802 (2018)

\bibitem{VoD19}
T.~Vockerodt, A.~Diaz-Torres, Phys. Rev. C \textbf{100}, 034606 (2019)

\bibitem{GSB74}
D.A. Goldberg, S.M. Smith, G.F. Burdzik, Phys. Rev. C \textbf{10}, 1362 (1974)

\bibitem{HNV82}
M.S. Hussein, H.M. Nussenzveig, A.C.C. Villari, J.L. Cardoso~Jr, Phys. Lett. B
  \textbf{114}, 1 (1982)

\bibitem{HuS84a}
M.S. Hussein, M.C.B.S. Salvadori, Phys Letters B \textbf{138}, 249 (1984)

\bibitem{HaR15}
K.~Hagino, N.~Rowley, EPJ Web of Conferences \textbf{86}, 00014 (2015)

\bibitem{BFG75}
J.B. Ball, C.B. Fulmer, E.E. Gross, D.C. Halbert, M.~L.~and.Hensley, C.A.
  Ludemann, M.J. Saltmars, G.R. Satchler, Nucl. Phys. A \textbf{252}, 208
  (1975)

\end{thebibliography}
\end{document}